\documentclass[12pt, preprint]{aastex}
\usepackage{amsmath}

\shorttitle{Cyclic evolution of radio pulsars}
\shortauthors{Beskin et al.}

\begin{document}

\title{Cyclic evolution of radio pulsars on the time scale of hundreds
of years}

\author{G. Beskin\altaffilmark{1, 3}, A. Biryukov\altaffilmark{2} and
S.Karpov\altaffilmark{1, 3} }

\altaffiltext{1}{Special Astrophysical Observatory of Russian Academy of
Sciences, Russia}
\altaffiltext{2}{Sternberg Astronomical Institute of Moscow State University,
Moscow, Russia}
\altaffiltext{3}{Isaac Newton Institute of Chile, SAO branch, Russia}

\begin{abstract}
The recent massive measurements of  pulsar frequency second derivatives have shown
that they are 100-1000 times larger than expected for standard pulsar
slowdown low. Moreover, the second derivatives as well as braking indices
are even negative for about half of pulsars. We explain these paradoxical results on the
basis of the statistical analysis of the rotational parameters
$\nu$, $\dot \nu$ and $\ddot \nu$ of the subset of 295 pulsars taken mostly from
the ATNF database. We have found strong correlation of $\ddot \nu$ and
$\dot \nu$  either for $\ddot\nu > 0$ (correlation coefficient $r\approx0.9$)
and $\ddot\nu < 0$ ($r\approx0.85$), and of $\nu$ and $\dot\nu$
($r\approx0.7$). We interpret these dependencies as evolutionary ones due to
$\dot\nu$ being nearly proportional to characteristic age $\tau_{ch}$. The derived
statistical relations as well as "anomalous" values of $\ddot\nu$ are well
explained in the framework of the simple model of cyclic evolution of the
rotational frequency of the pulsars. It combines the secular change
of $\nu_{tr}(t)$, $\dot\nu_{tr}(t)$ and $\ddot\nu_{tr}(t)$ according to the 
power law with $n\approx5$ and harmonic oscillations of 100--1000 years
period with an amplitude from $10^{-3}$ Hz for young pulsars to $10^{-10}$ Hz
for elder ones. The physical nature of these cyclic variations of the
rotational frequency may be similar to the well-known red timing noise,
however, with much larger characteristic time scale.
\end{abstract}


\section{Introduction}

The slowdown of radio pulsars is due to the electromagnetic emission energy
losses. According to this "classical" approach, their rotational frequencies
$\nu$ evolve following the slowdown law $\dot\nu=-K\nu^n$, where $K$ is the
positive constant that depends on the magnetic dipole moment and the moment
of inertia of the neutron star, $n$ is the braking index. The latter can be
determined observationally from measurements of $\nu$, $\dot \nu$ and $\ddot
\nu$ as $n={\nu \ddot \nu}/{\dot \nu^2}$. For pure dipole structure of pulsar
magnetosphere $n=3$, the pulsar wind decreases this value down to $n=1$,
while for multipole magnetic field it is up to $n=5$ \citep{man77, man85,
bla88}. For such law of pulsar evolution the second derivatives of their
rotational frequencies (and therefore their braking indices) can't be
measured for objects with the characteristic age $\tau_{ch} =
-\nu / 2 \dot\nu > 10^6$ years as their 
$\ddot \nu < 10^{-30}$ s$^{-3}$, that is much lower than the current
detection limit \citep{liv05}. At the same time the massive measurements of
pulsar frequency second derivatives $\ddot \nu$ have shown that their values
are much larger than expected for standard spin-down law and are even
negative for about half of all pulsars. The corresponding braking
indices are from $-10^6$ to $10^6$ \citep{d'a93, chu03, hob04}.

These strange results are usually interpreted as an influence of a red timing
noise, which manifests itself as irregular variations of pulse arrival
times relative to the secular frequency evolution \citep{gro75, cor80, cor85},
described through Taylor series expansion as
\begin{equation}
\nu(t)=\nu_0+\dot\nu_0(t-t_0)+\frac12\ddot\nu_0(t-t_0)^2+...
\end{equation}
where $t_0$, $\nu_0$, $\dot \nu_0$ and $\ddot \nu_0$ are initial time moment,
frequency and its derivatives at this moment. Historically, the secular
evolution of pulsar rotation rate has been considered as the linear trend of
the frequency behaviour, while quadratic term of the series (1) (with cubic
behaviour in pulse arrival times), measured as $\ddot \nu$, has been treated
as the signature of non-monotonic (on time scale of the observations)
processes -- red noise, post-glitch recovery, pulsar orbital motion and free
precession \citep{gro75, cor80, she96, sta00}. However, recently the increase
of the sensitivity of the receivers and time spans of the continuous
timing (from several to thirty years) made it possible to estimate the
higher-order coefficients of the series (1) \citep{cam00, sco03, hob04,
liv05}. The results of these studies show that $\ddot \nu$, while changing on
the time scales shorter than observation time spans, {\it on average}
characterizes the pulsar secular evolution rather than noise processes. The
latter, causing the variations mentioned earlier, worsens the accuracy of the
second derivative determination, while can't change its mean value
hundreds times \citep{cam00, hob04, liv05}.

In this Letter we report the detection of correlation of $\ddot\nu$ and
$\dot\nu$ either for $\ddot\nu > 0$ and $\ddot\nu < 0$ and correlation
of $\nu$ and $\dot\nu$. We show that these statistical relations as well
as "anomalous" values of $\ddot\nu$ can be explained by the combination
of secular decrease of pulsar frequencies with $n\approx 5$ and their cyclic
variations on the time scale of several hundreds of years.

\section{Statistical analysis of the ensemble of pulsars}

Our statistical analysis is based on the assumption that numerous
measurements of the pulsar frequency second derivatives reflect their secular
evolution on the time scale larger than the duration of observations, and
uses the parameters of nearly 300 pulsars.

\begin{figure}[t]
{\centering \resizebox*{1\columnwidth}{!}{\includegraphics[angle=270]
{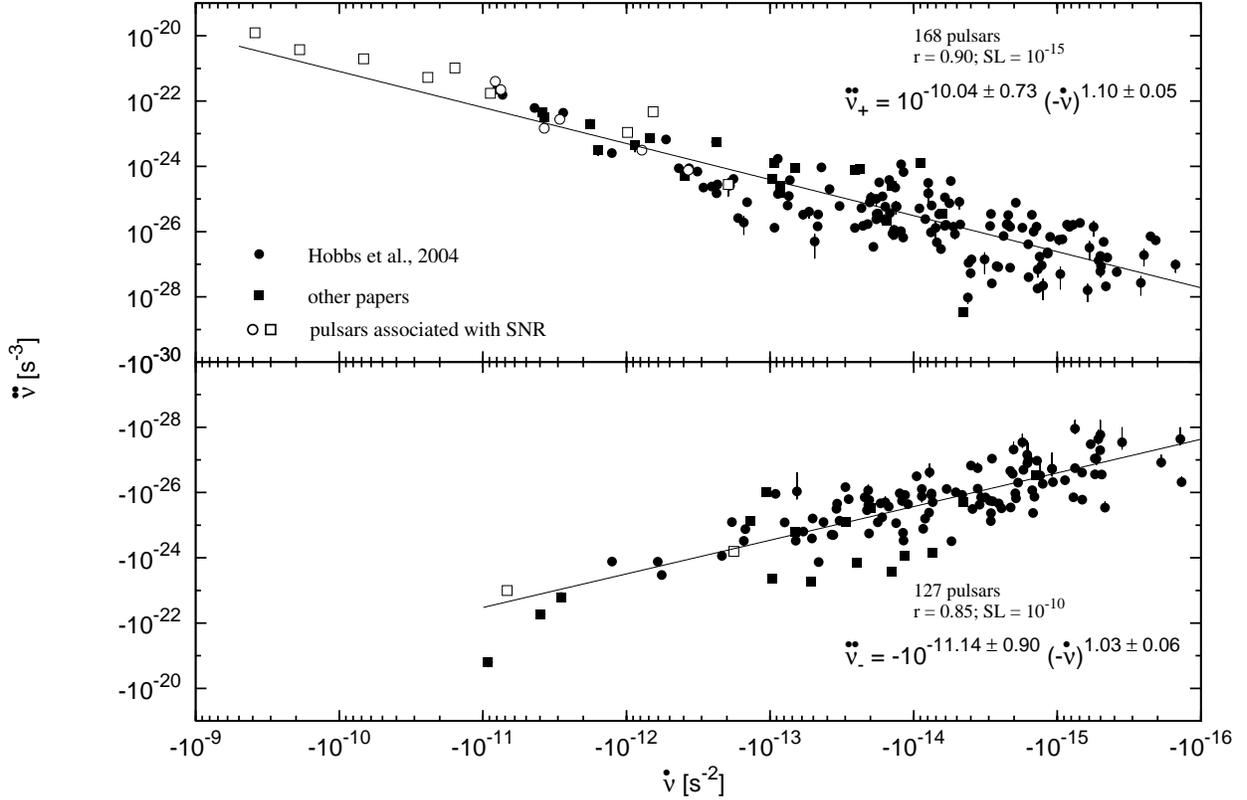}} \par}
\caption{The $\ddot \nu - \dot \nu$ diagram. The figure shows the pulsars
from the work \citep{hob04} as
circles, and the objects measured by other groups as squares. Empty symbols
 represent the pulsars associated
with the supernova remnants, and therefore -- relatively young ones.
Analytical fits for both positive and negative
branches are shown as solid lines.}
\end{figure}

From 389 objects of ATNF catalogue \citep{man05} with known $\ddot \nu$ we
have compiled a list of "ordinary" radio pulsars with $P>20$ ms 
and $\dot P
>10^{-17}$ s s$^{-1}$, excluding recycled, anomalous and binary pulsars, and
with relative accuracy of second derivative measurements better than 75\%.
It has been appended with 26 pulsars from other sources \citep{d'a93, chu03}.
The parameters of all pulsars have been plotted on the $\ddot \nu - \dot \nu$
diagram (Fig. 1). Here the circles are the objects with second derivatives
measured at the Jodrell Bank Observatory \citep{hob04}, squares -- the results
of other groups \citep{d'a93, chu03}, and empty symbols represent young
pulsars associated with supernova remnants.

The main result of the statistical analysis of these data is a significant
correlation of $\ddot \nu$ and $\dot \nu$, either for 168 objects with $\ddot
\nu > 0$ (correlation coefficient $r\approx0.90$) and for 127 objects with
$\ddot \nu < 0$ ($r\approx0.85$). Either groups follow the nearly linear laws,
however they are not exactly symmetric with respect to $\ddot \nu=0$. Also,
the young pulsars with $\dot \nu < -10^{-11}$ s$^{-2}$ and $\ddot \nu < 0$
are absent.

\begin{figure}[t]
{\centering \resizebox*{1\columnwidth}{!}{\includegraphics[angle=270]
{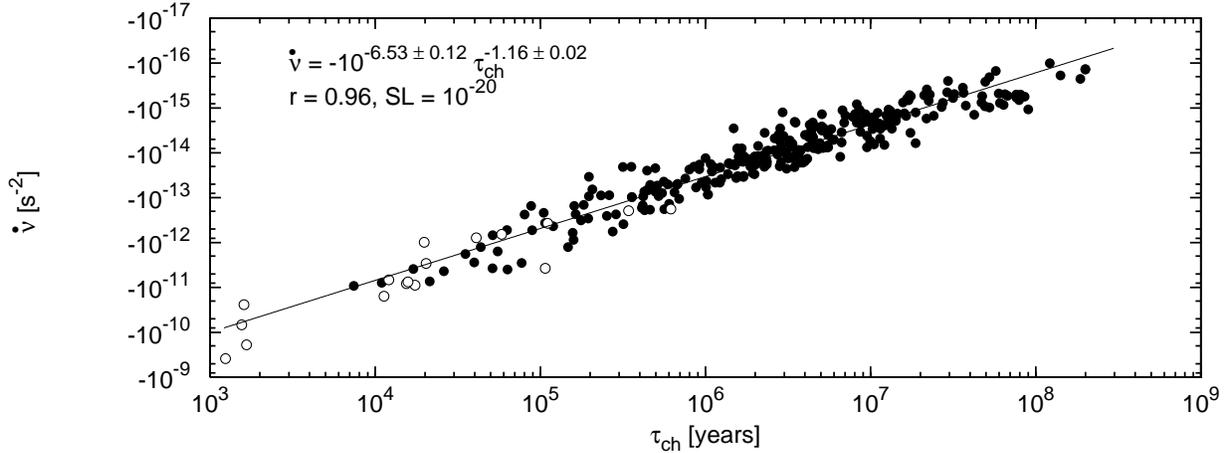}} \par}
\caption{The correlation of $\dot\nu$ and $\tau_{ch}$.}
\end{figure}

Figure 2 shows the dependence of $\dot \nu$ on the characteristic age
$\tau_{ch}$. Of course, due to practical proportionality of these parameters
there is a strong dependence of $\ddot \nu$ on $\tau_{ch}$ either for pulsars
with $\ddot \nu >0$ (positive branch, $r \approx 0.85$) and for $\ddot \nu <
0$ (negative branch, $r \approx 0.75$) (Fig. 3a). Also, we have found the
significant correlation of $n$ and $\tau_{ch}$ (Fig. 3b). Note some important
points of these results.

\begin{itemize}
\item{
Measurements of different observational groups show the same laws, so
the correlations discovered are not related to the peculiarities of the
receiving devices of a single radio telescope.
}
\item{
Young pulsars confidently associated with supernova remnants are
systematically shifted to the left in  Figure 1 and Figure 2 (empty symbols).
The order of their physical ages roughly corresponds to that of
characteristic ages. It means that any dependence on $\dot \nu$ or
$\tau_{ch}$ may be interpreted as the dependence on time.
}
\item{
Separate analysis of the whole set of parameters of pulsars with $\ddot \nu > 0$
and $\ddot \nu < 0$ (galactic positions, fluxes, luminosities, magnetic fields,
dispersion and rotation measures) shows the statistical identity of these
sub-groups.
}
\end{itemize}

So, the $\ddot\nu - \dot \nu$ diagram (Fig. 1) may be interpreted as an
evolutionary one. In other words, each pulsar during its evolution moves
along the branches of this diagram while increasing the value of its
$\dot \nu$ (which corresponds to the increase of its characteristic age).
However, there is an obvious contradiction, as for $\ddot \nu < 0$ (negative
branch), $\dot \nu$, being negative, may only decrease with time, so the
motion along the negative branch may be only backward! Certainly, this
contradiction is easily solved by assuming non-monotonic behaviour of
$\nu(t)$, which may have some cyclic (quasi-periodic) component along with
secular one. Figure 4 shows the example of such cyclic behaviour of a pulsar.
The characteristic time scale $T$ of such cycles must be much shorter than
the pulsar life time and at the same time much larger than the time scale of
the observations. Each pulsar during its evolution repeatedly changes the
sign of $\ddot \nu$ and spends roughly half of its life time on each branch.
The asymmetry of the branches reflects the positive secular behaviour of
$\ddot \nu_{tr}(t)$, and so, secular increase of $\dot\nu(t)$. Systematic
decrease of branches separation reflects the decrease of the oscillations
amplitude and/or the increase of its characteristic time scale. Any
non-monotonic variations of $\nu(t)$, like post-glitch recovery, standard red
noise or orbital motion, will manifest itself in a similar way on the
$\ddot\nu - \dot \nu$ diagram and lead to extremely high values of
$\ddot \nu$ \citep{gro75, cor80, she96, sta00}. However, their characteristic
time scales vary from weeks to years, and they are detected immediately. In
our case we have processes on time scales of hundreds of years, and their
study is possible only by means of statistical methods, assuming that the
ensemble of pulsars of various age models the behaviour of a single pulsar
during its evolution.

\begin{figure}[t]
{\centering \resizebox*{1\columnwidth}{!}{
\includegraphics[angle=270]{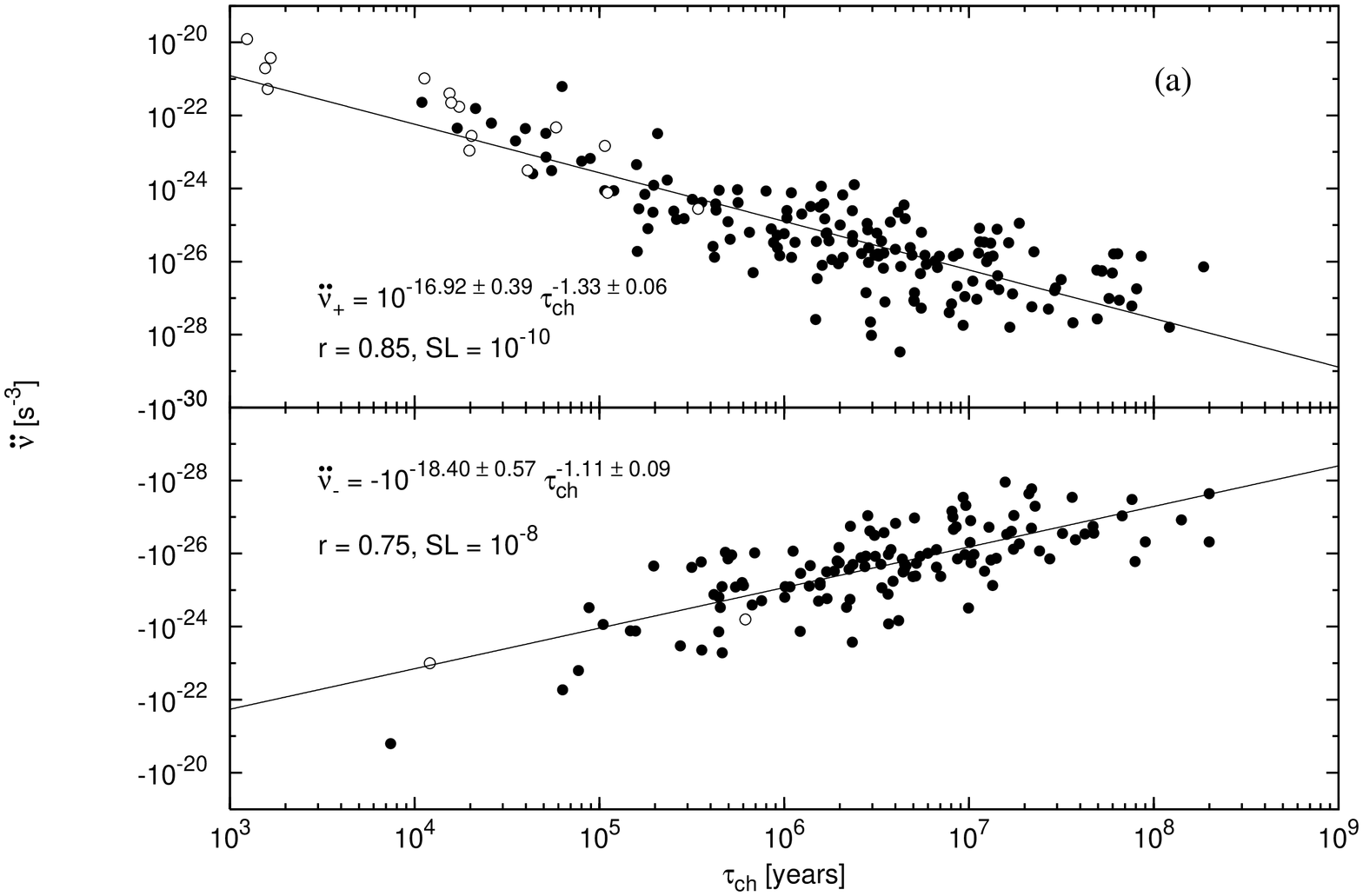}
}\par}
{\centering \resizebox*{1\columnwidth}{!}{
\includegraphics[angle=270]{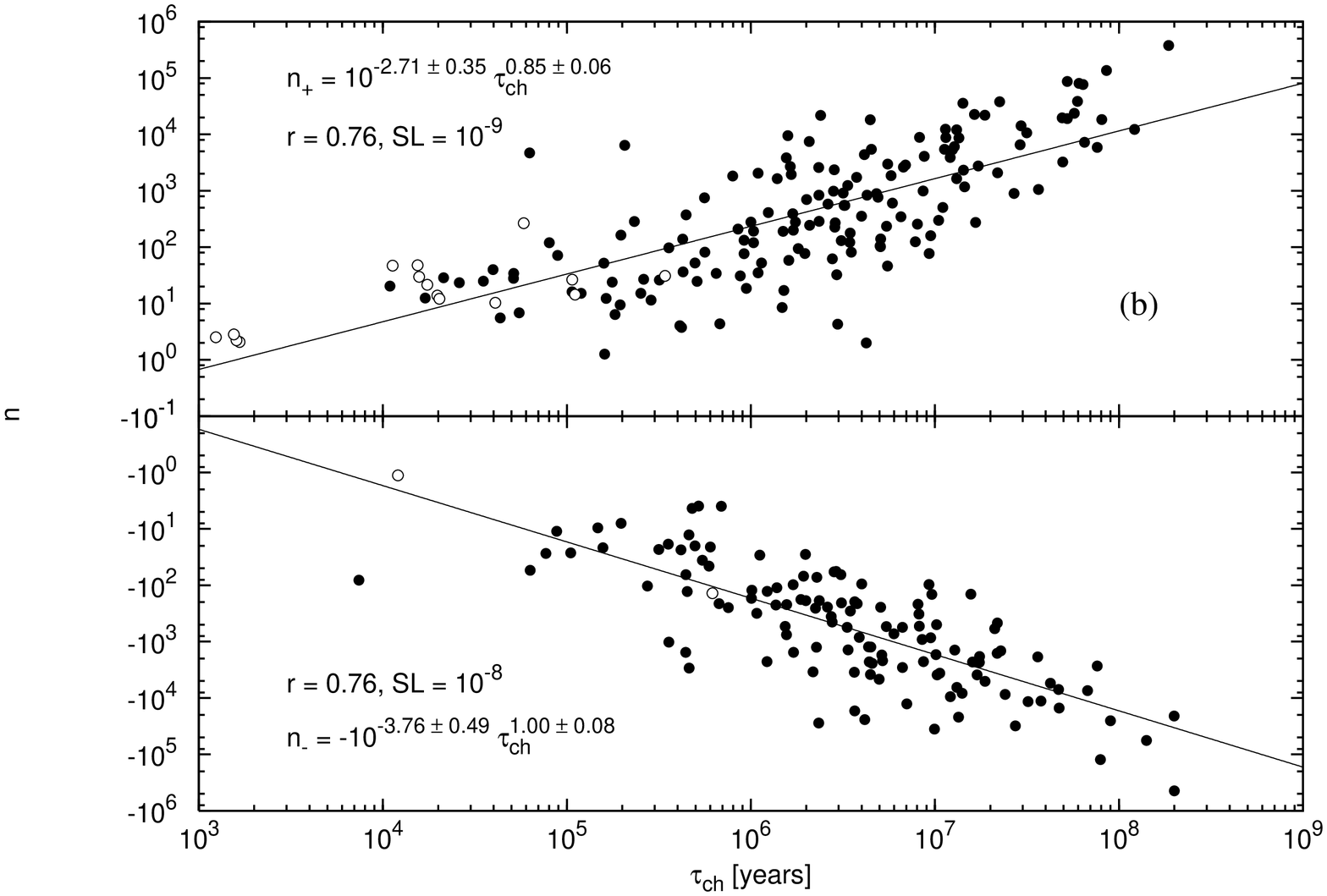}
}\par}
\caption{(a) The correlation of $\ddot\nu$ and $\tau_{ch}$. Empty symbols
represent the young pulsars associated with the supernova remnants.
It is clear that the pulsar characteristic age roughly corresponds
to the real one. (b) The correlation of $n$ and $\tau_{ch}$.}
\end{figure}

\begin{figure}[t]
{\centering \resizebox*{1\columnwidth}{!}{\includegraphics[angle=270]
{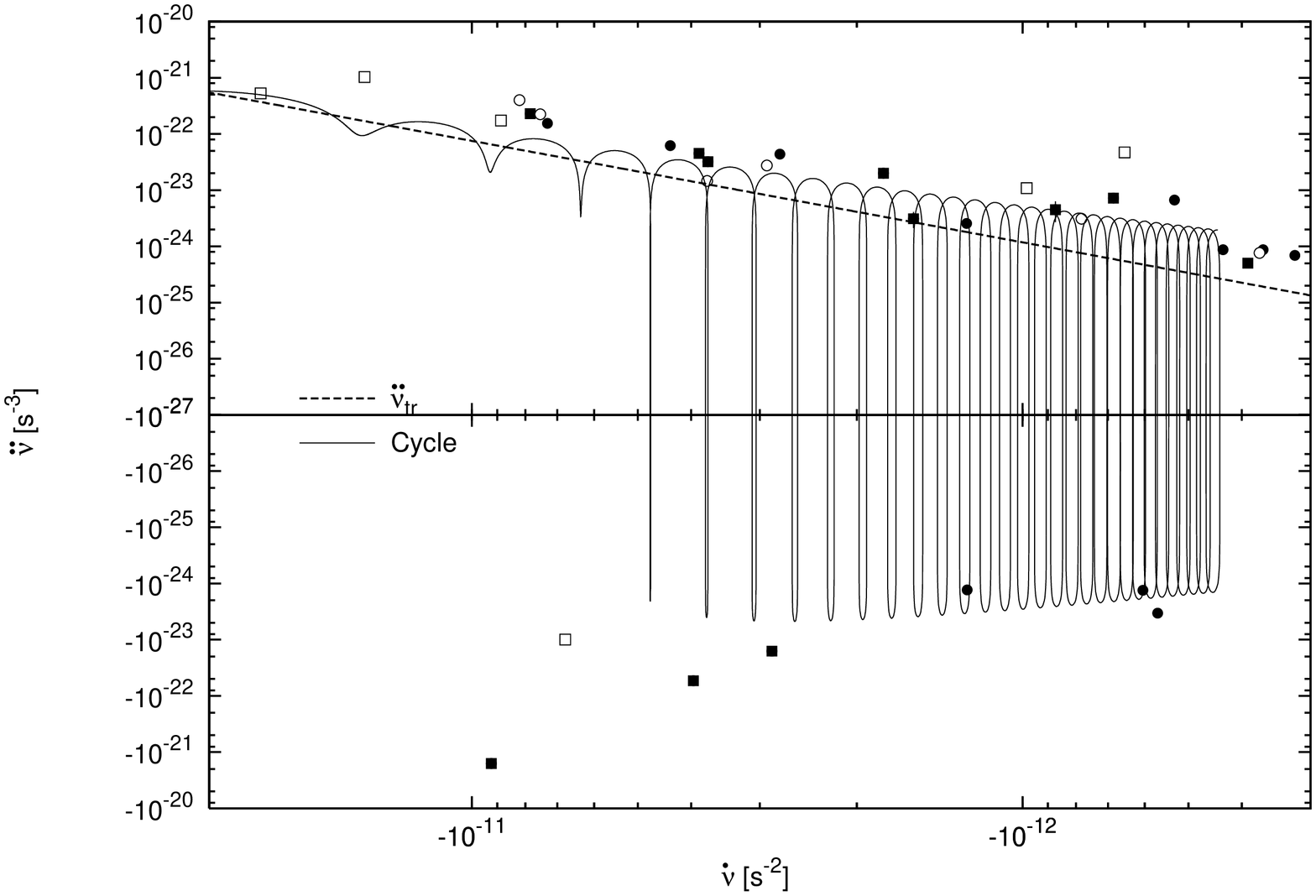}} \par}
\caption{The sample of pulsar cyclic behaviour on the $\ddot \nu - \dot \nu$
diagram. The solid line shows the evolution of a pulsar according to the
simple harmonic model presented in (2) -- (7) with the oscillation period
$T=1000$ years. The dashed line shows the corresponding secular evolution
term $\ddot\nu_{tr}(t)$, and the oscillations are around it, symmetric in
linear scale. For $\dot\nu<-7\cdot10^{-12}$ s$^{-2}$, the amplitude of the
oscillations is smaller than the secular term, thus $\ddot\nu$ is always
positive, for larger values it changes its sign repeatedly, spending
approximately half of the time in the negative region.}
\end{figure}

Note that the correlations between $|\ddot P|$ and $\dot P$, similar to these
discovered here (but without distinguishing between objects with $\ddot \nu >
0$ and $\ddot \nu < 0$), were discussed earlier \citep{cor85, arz94, lyn99}
as an evidence of higher timing noise of young pulsars.

\section{The simplest model of pulsar cyclic evolution}

Cyclic variations of the pulsar rotational frequency may be complicated --
quasi-periodic, multi-harmonic, stochastic. As a rough initial approximation,
we use a simple model of harmonic oscillations superimposed on secular
evolution. In this case,
\begin{equation}
\nu(t)=\nu_{tr}(t) + A(t)\sin\left[{\Omega(t)t + \phi_0}\right] ,
\end{equation}
where $A(t)$ is the amplitude of periodic frequency variations, 
$\Omega(t)=2\pi / T(t)$ -- its frequency, and $\nu_{tr}(t)$ is the
secular evolution term. This leads to
\begin{equation}
\dot \nu(t)=\dot \nu_{tr}(t) + A(t)\Omega(t)\cos\left[{\Omega(t)t +
\phi_0}\right] ,
\end{equation}
\begin{equation}
\ddot \nu(t)=\ddot \nu_{tr}(t) - A(t)\Omega^2(t)\sin\left[{\Omega(t)t +
\phi_0}\right] ,
\end{equation}
where we neglected the terms with $A(t)$ and $\Omega(t)$ derivatives due
to its relative smallness. From (3) and (4) it is easy to get the relation
between $\ddot \nu(t)$ and $\dot \nu(t)$ in the cycloidal form
$[\ddot \nu(t)-\ddot\nu_{tr}(t)]^2=\Omega^4(t)A^4(t)- \Omega^2(t)[\dot\nu(t)-
\dot\nu_{tr}(t)]^2$, which is plotted in Figure 4.

By marking with "+" and "-" subscripts the values related to the positive
and negative branches on the $\ddot\nu - \dot\nu$ diagram (Fig. 1),
correspondingly, and pointing to the amplitude values of $\nu$ and $\ddot
\nu$ (i.e. assuming $\sin{\left[\Omega(t)t + \phi_0\right]} = \mp 1$), we get
\begin{equation}
\nu_{\pm}(t) = \nu_{tr}(t) \mp A(t)
\end{equation}
\begin{equation}
\dot \nu_{\pm}(t)=\dot \nu_{tr}(t)
\end{equation}
\begin{equation}
\ddot \nu_{\pm}(t) = \ddot \nu_{tr}(t) \pm A(t)\Omega^2(t).
\end{equation}
So, due to (6), the branches of $\ddot\nu - \dot\nu$ diagram describe the
dependence of pulsar frequency second derivative $\ddot\nu$ on the secular
behaviour of its first derivative $\dot \nu_{tr}$, which corresponds to the
dependence on time.

Relations (2) -- (7) describe the apparent evolution of some "average"
pulsar. Of course, in the ensemble of all pulsars the scatter of points on
the $\ddot\nu - \dot\nu$ diagram reaches 4 orders of magnitude due to
distribution of the neutron star parameters and initial conditions. Due to
this fact it is impossible to determine $\ddot \nu_{tr}(t)$ directly as the
half-sum of $\ddot \nu_{+}(t)$ and $\ddot \nu_{-}(t)$ from (7) and Fig. 1, as
it requires the subtraction of two close values with significant errors.

First of all we compute the behaviour of $\nu_{tr}(t)$ (or, $\nu_{tr}(\dot
\nu_{tr})$) by plotting the studied pulsar group on the $\dot \nu - \nu$ diagram
(Fig. 5). The objects with $\ddot \nu > 0$ and $\ddot \nu < 0$ are marked as
filled and empty circles, correspondingly. It is easily seen that the
behaviour of these two sub-groups is the same, i.e. $A(t)$ according to (5)
is significantly smaller than the intrinsic scatter of $\nu(\dot \nu)$.
However, a strong correlation between $\nu_{tr}$ and $\dot\nu_{tr}$
($r\approx0.7)$ is seen, and
\begin{equation}
\dot\nu_{tr}=-C\nu_{tr}^n,
\end{equation}
where $C=10^{-15.26\pm1.38}$ and $n=5.13\pm0.34$.
So, the secular evolution of the "average" pulsar is according to the
"standard" spin-down law with $n\approx5$! This value may suggest the
importance of multipole components of pulsar magnetic field, or the deviation
of the angle between pulsar rotational and dipole axes from ${\pi}/{2}$
\citep{man77, con05}. From (8) we may easily get the relation between
$\ddot\nu_{tr}$ and $\dot\nu_{tr}$
\begin{equation}
\ddot\nu_{tr} (\dot\nu_{tr}) = nC^{\frac1n}(-\dot\nu_{tr})^{2-\frac1n},
\end{equation}
which is shown in Figure 6 as a thick dashed line. The amplitude of the
$\ddot \nu(t)$ oscillations, $A_{\ddot \nu}(t)$, may be easily computed
from (7) using fits for $\ddot\nu_{+}$ and $\ddot\nu_{-}$ in Figure 1,
\begin{equation}
A_{\ddot \nu}(t) = A(t)\Omega^2(t)=\frac12\left[\ddot \nu_{+}(t) - \ddot
\nu_{-}(t)\right]
\end{equation}

\begin{figure}[t]
{\centering \resizebox*{1\columnwidth}{!}{\includegraphics[angle=270]
{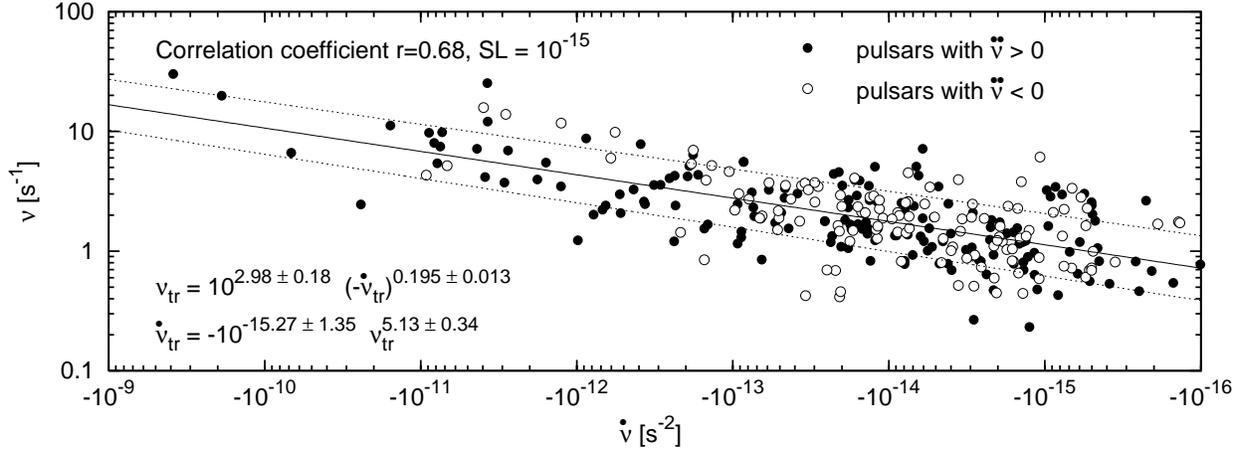}} \par}
\caption{The $\dot \nu - \nu$ diagram for the pulsars with the measured
second derivative. The filled symbols are objects with positive $\ddot\nu$,
the empty ones -- with negative. The behaviour of both subsets is the same.
The solid line represents the best fit, corresponding to the $n\approx5$
braking index, the dotted ones -- 1-$\sigma$ range.}
\end{figure}

\begin{figure}[t]
{\centering \resizebox*{1\columnwidth}{!}{\includegraphics[angle=270]
{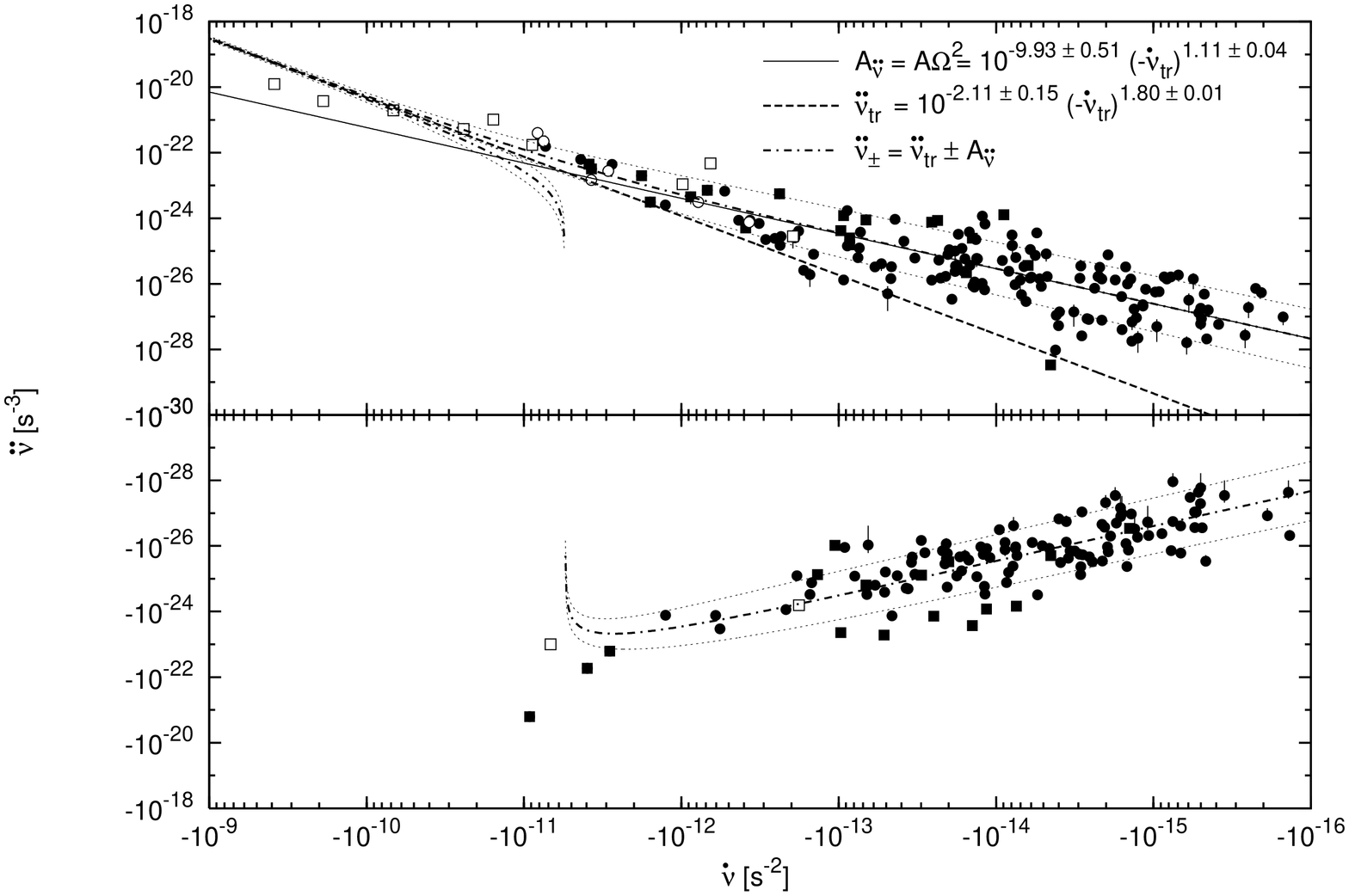}} \par}
\caption{The $\ddot \nu - \dot \nu$ diagram with the simple harmonic
oscillations model. The solid line is the amplitude $A_{\ddot\nu}$
of the frequency second derivative variations  according to (10),
the dashed -- the secular term $\ddot\nu_{tr}$ from (9), and the dot-dashed
lines are the envelopes of the oscillations $\ddot\nu_{tr}\pm A_{\ddot\nu}$
with 1-$\sigma$ ranges (dotted lines). Each pulsar spends the majority of
its life time at or very near the envelopes.}
\end{figure}

The result is shown in Figure 6 as a solid line. The amplitude $A_{\ddot\nu}$
is decreasing almost linearly with the increase of $\dot \nu_{tr}$, i.e. life
time. This may be related to the decrease of either $A(t)$ or $\Omega(t)$.

From Figure 6 it is seen that our simple model formulated in (2) -- (7)
describes the observational data rather well, at least for the pulsars with
$\dot\nu>-10^{-11}$ s$^{-2}$, i.e. with $\tau_{ch}>10^4$ years. For
$\dot\nu<-7\cdot10^{-12}$ s$^{-2}$, the $A_{\ddot\nu}$ is smaller than
$\ddot\nu_{tr}$, and during the oscillations $\ddot\nu>0$ always, that
explains the absence of pulsars of the negative branch of $\ddot\nu-\dot\nu$
diagram in this region. The situation changes dramatically for
$\dot\nu>10^{-11}$ s$^{-2}$, i.e. $\tau_{ch} > 10^4$ years. Here
$A_{\ddot\nu}$ becomes larger than $\ddot\nu_{tr}$ and pulsars with
$\ddot\nu < 0$ and $n < 0$ appear. In other words, the cyclic behaviour
according to the model described by (2) -- (7) begins to manifest itself in
the observing quantities, -- $\ddot\nu$ and $n$ begins to deviate
significantly from the predictions of the "standard" spin-down law evolution.

On the basis of these assumptions it is possible to get a rough estimate
of the maximal period of the cyclic behaviour as the several times of the
time scale of the vanishing of $\ddot\nu$ for the young pulsars with the
measured $\dddot\nu$ as $T_{max}\approx-4{\ddot\nu}/{\dddot\nu}$. For {\it
PSR B1509-58} (with $\ddot\nu\approx2\cdot10^{-21}$ s$^{-3}$ and
$\ddot\nu\approx-1.3\cdot10^{-31}$ s$^{-4}$) \citep{liv05} $T_{max}\approx
2\cdot10^{3}$ years ($\Omega_{min}\approx10^{-10}$ s$^{-1}$). For {\it Crab}
(with $\ddot\nu\approx1.2\cdot10^{-20}$ s$^{-3}$ and $\ddot\nu\approx-
1.4\cdot10^{-30}$ s$^{-4}$) \citep{sco03} $T_{max}\approx 10^{3}$ years
($\Omega_{min}\approx2\cdot10^{-10}$ s$^{-1}$). On the other hand, the
estimations of $\Omega$ may be done using the relations (2) -- (4), for the
pulsars with several measurements of rotational parameters at different
epochs with enough separation. So, the lower limit is $\Omega_{min}\approx
\sqrt{\left|{\Delta\ddot\nu}/{\Delta\nu}\right|}$. From the results of
observations of the old ($\tau \propto 10^6$ years) pulsars, {\it B0823+26},
{\it B1706-16}, {\it B1749-28}  and {\it B2021+51} with 7 to 10 years
separation, we have obtained the same estimate of $T_{max}\approx10^3 \div
2\cdot 10^3$ years. Finally, the period of cyclic oscillations is likely to
be in the range of 100 -- 1000 years ($2\cdot10^{-10} < \Omega < 2\cdot
10^{-9}$ s$^{-1}$). Here we have suggested that the minimal cycle period has
to be several times the pulsar observation period -- 30 years.

Taking into account that $A_{\ddot \nu} = \Omega^2A$ is changing several
orders of magnitude, our initial assumption of relative constancy of $\Omega$
and significant decrease of $A$ during the evolution is somehow justified.
Assuming $\Omega\approx10^{-9}$ s$^{-1}$, we get for the amplitude of
frequency oscillations $A={A_{\ddot \nu}}/{\Omega^2}$ values from $4\cdot
10^{-4}$ s$^{-1}$ for the youngest pulsars till $10^{-10}$ s$^{-1}$ for the
eldest.

\section{Discussion and conclusions}

The physical reasons of the discussed cyclic variations of the pulsar
rotational frequency may be similar to the ones of the well-known red timing
noise. In the first place, the free precession \citep{sta00} is the periodic
effect most suitable for the proposed harmonic model. Note that the
interaction of the superfluid core and the crust of the neutron star does not
depress its development \citep{alp05}. Several processes had been proposed
for the explanations of the red noise \citep{cor80, cor81, che87} -- from the
collective effects in the superfluid core till the electric current fluctuations in
the pulsar magnetosphere. The possibility of its utilization for the
explanation of the frequency variations on the time scale of hundreds of
years is yet to be analyzed.

The argument towards the similarity between the discussed variations and
the red timing noise is the coincidence of the red noise $\ddot\nu$ amplitude
extrapolated according to its power spectrum slope \citep{bay99} to the time
scale of hundreds of years, with $A_{\ddot\nu}$ in our model for the same
$\dot\nu$, i.e. the same ages (see Fig. 6).

Finally, note that the real evolution of the pulsar rotational frequencies
may be significantly more complicated than the proposed simple model. The
variations may be non-harmonic, their amplitudes, phases and frequencies may
change randomly. The principal point is that all the pulsars evolve this
way, and the time scale of such variations must significantly exceed several
tens of years. That explains the anomalous values of the observed $\ddot\nu$
and braking indices.

\acknowledgments

This work has been supported by the Russian Foundation for Basic
Research (grant No 04-02-17555), Russian Academy of Sciences (program
"Evolution of Stars and Galaxies"), and by the Russian Science Support
Foundation.


\begin{thebibliography}{1}

\bibitem[Alpar(2005)]{alp05}
Alpar, M. A. 2005, preprint (astro-ph/0505073)

\bibitem[Arzoumanian et al.(1994)]{arz94}
Arzoumanian, Z., Nice, D. J., Taylor, J. H., \& Thorsett, S. E. 1994,
\apj, 422, 671

\bibitem[Baykal et al.(1999)]{bay99}
Baykal, A., Alpar, M.A., Boynton, P.E. \& Deeter, J.E. 1999,
\mnras, 306, 207

\bibitem[Blandford \& Romani(1988)]{bla88}
Blandford, R.D \& Romani R.W. 1988, \mnras, 234, 57P

\bibitem[Camilo et al.(2000)]{cam00}
Camilo, F., Kaspi, V. M., Lyne, A. G., Manchester, R. N., Bell, J. F.,
D'Amico, N., McKay, N. P. \& Crawford, F. 2000, \apj, 541, 367

\bibitem[Cheng (1987)]{che87}
Cheng, K.S. 1987, \apj, 321, 799

\bibitem[Chukwude(2003)]{chu03}
Chukwude, A.E. 2003, \aap, 406, 667

\bibitem[Contopoulus \& Spitkovsky(2005)]{con05}
Contopoulus, I., \& Spitkovsky, A. 2005, preprint (astro-ph/0512002)

\bibitem[Cordes \& Helfand(1980)]{cor80}
Cordes, J. M. \& Helfand, D. J. 1980, \apj, 239, 640

\bibitem[Cordes \& Greenstein(1981)]{cor81}
Cordes, J. M. \& Greenstein, G. 1981, \apj, 245, 1060

\bibitem[Cordes \& Downs(1985)]{cor85}
Cordes, J. M. \& Downs, G. S. 1985, \apjs, 59, 343

\bibitem[D'Alessandro et al.(1993)]{d'a93}
D'Alessandro, F., McCulloch, P. M., King, E. A., Hamilton, P. A. \& McConnell,
D. 1993, \mnras, 261, 883

\bibitem[Groth(1975)]{gro75}
Groth, E.J. 1975, \apj, 529, 431

\bibitem[Hobbs et al.(2004)]{hob04}
Hobbs, G. B., Lyne, A. G., Kramer, M., Martin, C. E. \& Jordan, C.
2004, \mnras, 353, 1311

\bibitem[Livingstone et al.(2005)]{liv05}
Livingstone, M. A., Kaspi, V. M. Gavriil, F. P., \& Manchester, R. N. 2005,
\apj, 619, 1046

\bibitem[Lyne(1999)]{lyn99}
Lyne, A. 1999, in Pulsar Timing, General Relativity and the Internal
Structure of Neutron Stars, ed. Z. Arzoumanian, F. Van der Hooft, \&
E. P. J. van den Heuvel (Amsterdam: Koninklijke Nederlandse Akademie van
Wetenschappen), 141

\bibitem[Manchester \& Taylor(1977)]{man77}
Manchester, R. N. \& Taylor, J. H. 1977, Pulsars (San Francisco: Freeman)

\bibitem[Manchester, Durdin \& Newton(1985)]{man85}
Manchester, R. N., Durdin, J. M., \& Newton, L. M. 1985, Nature, 313, 374

\bibitem[Manchester et al.(2005)]{man05}
Manchester, R. N., Hobbs, G. B., Teoh, A. \& Hobbs, M. 2005, \aj, 129,
1993

\bibitem[Scott, Finger \& Wilson(2003)]{sco03}
Scott, D., Finger, M. \& Wilson, C. 2003, \mnras, 344, 412

\bibitem [Shemar \& Lyne(1996)]{she96}
Shemar, A. L. \& Lyne, A. G. 1996, \mnras, 282, 677

\bibitem[Stairs, Lyne \& Shemar(2000)]{sta00}
Stairs, I. H., Lyne, A. G. \& Shemar, A. L. 2000, Nature, 406, 484

\end{thebibliography}
\end{document}